\documentstyle[aps,graphicx,epsf]{revtex}\tighten
\def \BE {\begin{equation}}
\def \EE {\end{equation}}
\def \nt {\tau}

\newcommand{\edth}{\mathop{\hbox{\usefont{T1}{cmr}{medium}{n} \dh \hspace{-0.5mm}}}\nolimits}
\newcommand{\thorn}{\mathop{\hbox{\usefont{T1}{cmr}{medium}{n} \th \hspace{-0.5mm}}}\nolimits}

\begin{document}
\title{VSI$_i$ spacetimes and the $\varepsilon$-property}

\author{N. Pelavas\dag, A. Coley\dag, R. Milson\dag, V. Pravda\ddag, A. Pravdov\'{a}\ddag}
\address{\dag\ Department of Mathematics and Statistics, Dalhousie University, Halifax, Nova Scotia}
\address{\ddag Mathematical Institute, Academy of Sciences, \v Zitn\' a 25, 115 67 Prague 1, Czech Republic.}

\maketitle

\begin{abstract}
We investigate Lorentzian spacetimes where all zeroth and first order curvature invariants vanish and discuss how this
class differs from the one where all curvature invariants vanish (VSI).  We show that for VSI spacetimes all components
of the Riemann tensor and its derivatives up to some fixed order can be made arbitrarily small.  We discuss this in
more detail by way of examples.
\end{abstract}

\section{Introduction}

Recently it was proven that in 4 dimensional Lorentzian spacetimes all of the~scalar invariants constructed from
the~Riemann tensor and its covariant derivatives are zero if and only if the~spacetime is of Petrov (P)-type III, N or
O, all eigenvalues of the~Ricci tensor are zero and hence of Pleba\'{n}ski-Petrov (PP)-type N or O \cite{VSI} and the
common multiple null eigenvector of the Weyl and Ricci tensors is geodesic, shear-free, expansion-free and twist-free;
let us refer to these spacetimes as vanishing scalar invariant (VSI) spacetimes. VSI spacetimes include the well-known
pp-wave spacetimes \cite{jordan}.

Since all of the~scalar curvature invariants vanish, all VSI spacetimes are exact solutions of higher-order Lagrangian
based theories (in which the action is given by higher order scalar corrections to the usual general relativistic
action based on the Ricci scalar). It has subsequently been argued that, as in the case of pp-waves, VSI spacetimes are
exact solutions in string theory \cite{amati,HS,COLEY}, when supported by appropriate  bosonic massless fields of the
string (such as, for example, a dilaton and an antisymmetric massless field). Solutions of classical field equations
for which the~counter terms required to regularize quantum fluctuations vanish (i.e., they suffer no quantum
corrections to all loop orders) are also of importance because they offer insights into the~behaviour of the~full
quantum theory \cite{GG}.

In particular fundamental field theories
only certain specific types of higher order corrections occur (cf. \cite{cfmp,GSW,russot}),
and so for a spacetime to
be a solution of a particular field theory to all orders, with a specific
effective action containing only certain higher order correction terms, it may not be necessary
for {\em all} curvature invariants to vanish.
Consequently it is also of interest to determine the set of spacetimes for which (only)
the zeroth order curvature invariants vanish (i.e., algebraic scalar invariants
constructed from the Riemann tensor), denoted VSI$_0$,
those spacetimes for which (only)
the zeroth {\em and} first order curvature invariants vanish (i.e., scalar invariants
constructed from the Riemann tensor and its first covariant derivative), denoted VSI$_1$,
and so on. In fact, it was proven in \cite{VSI} that if all of
the zeroth, first and second order curvature invariants vanish, then necessarily
all scalar curvature invariants vanish; so that VSI$_2$
is equivalent to the set of VSI spacetimes.

Let us first recall some properties of VSI spacetimes. Utilizing a complex null tetrad in the Newman-Penrose (NP)
formalism it was shown that for P-types III and N the repeated null vector of the Weyl tensor $\ell^\alpha$ is
geodesic, shear-free, expansion-free and twist-free (and the NP coefficients $\kappa$, $\sigma$, and $\rho$ are
consequently zero), and the Ricci tensor has the form
\begin{equation}
 R_{\alpha \beta} = - 2\Phi_{22} \ell_{\alpha}
 \ell_{\beta} + 4 \Phi_{21}  \ell_{(\alpha} m_{\beta)}
+ 4 \Phi_{12}  \ell_{(\alpha} {\bar m}_{\beta)}, \label{Ricci}
\end{equation}
in terms of the non-zero Ricci components $\Phi_{ij}$. For P-type O, the~Weyl tensor vanishes and so it suffices that
the~Ricci tensor has the~form (\ref{Ricci}). All of these spacetimes belong to Kundt's class \cite{kundt}, and the
metrics for all VSI spacetimes are displayed in  \cite{VSI}. The generalized pp-wave solutions are of P-type N, PP-type
O (so that the~Ricci tensor has the~form of null radiation) with $\nt=0$, and admit a covariantly constant null vector
field \cite{jordan}. The~Ricci tensor (\ref{Ricci})  has four vanishing eigenvalues, and the PP-type is N    for
$\Phi_{12}\not= 0$ or O for $\Phi_{12}=0$. It is known that the energy conditions are violated in the PP-type N models
\cite{kramer} and hence attention is usually concentrated on the more physically interesting  PP-type O case, which in
the  non-vacuum case corresponds to pure radiation.

It is well known that the necessary and sufficient conditions for spacetimes for which the zeroth order algebraic
scalar curvature invariants vanish (VSI$_0$) are of P-type III, N or O and PP-type  N or O. Moreover, the repeated
principal null direction of Weyl must be aligned with an eigenvector of the Ricci tensor. The last condition follows
from the vanishing of the mixed invariants (see \S 3.1 of \cite{VSI}). Next we determine the VSI$_1$ spacetimes.

\section{VSI$_1$}
We begin by assuming VSI$_0$ and determine the conditions which imply VSI$_1$.  From the Bianchi identities it follows
for VSI$_0$ that $\kappa=0$. The invariants used here are all constructed from spinors that are symmetrized before and
after contractions. Since contractions are always performed with symmetrized spinors we need only give the number of
indices contracted between any two spinors. In particular, we shall make use of the following invariant,
$I_1\equiv(\nabla\Psi)^2(\nabla\overline{\Psi})^2$.
Here $(\nabla\Psi)^2$ is used to indicate the contraction over four indices of two copies of
$\nabla_{(A\dot{A}}\Psi_{BCDE)}$.  The result is then symmetrized and contracted with its conjugate to give
$I_1$. 

\vspace{0.5cm}

\noindent (a) \textit{Petrov type III.}

Using $\Psi_3 \neq 0$ with PP-type N or O, we have from the Bianchi identities that $\sigma\Psi_3 = \rho\Phi_{12}$ and
$\kappa = 0$. Applying $\kappa = 0$ throughout, we find that two of the Bianchi identities yield the following relation

\begin{equation}
D\Psi_3 = \rho\Phi_{21}+\overline{\sigma}\Phi_{12}+2(\rho-\varepsilon)\Psi_3.  \label{DPsi3}
\end{equation}

\noindent Computing $I_1$ and using (\ref{DPsi3}), we obtain

\begin{equation}
I_1 = \frac{576}{625}\left[81(\sigma\overline{\sigma}\Psi_3\overline{\Psi}_3)^2 +
\sigma\overline{\sigma}\Psi_3\overline{\Psi}_3 X\overline{X} + (X\overline{X})^2 \right],   \label{I1P3}
\end{equation}

\noindent where $X = \rho\Phi_{21}+\overline{\sigma}\Phi_{12}+5\rho\Psi_3$.

The vanishing of $I_1$ necessarily implies that $\sigma = 0$, thus from the Bianchi identities $\rho\Phi_{12}=0$.  If
$\rho = 0$ we get VSI. If $\Phi_{12}=0$ then (\ref{I1P3}) becomes $I_1 =
576(\rho\overline{\rho}\Psi_3\overline{\Psi}_3)^2$ which vanishes when $\rho=0$, giving VSI with PP-type O (null
radiation).

\vspace{0.5cm}

\noindent (b) \textit{Petrov type N.}

Using $\kappa = 0$ in the Bianchi identities we find that $\rho\Phi_{21}=-\overline{\sigma}\Phi_{12}$ and
$\rho\Phi_{12}=0$.  Therefore if $\Phi_{12} \neq 0$ then $\rho = 0$ implies that $\sigma = 0$, hence we recover VSI. If
$\Phi_{12}=0$ then two of the Bianchi identities combine to yield $\sigma\Psi_4 = \rho\Phi_{22}$.  The conditions
$\kappa = \Phi_{12} = 0$ and $\sigma\Psi_4 = \rho\Phi_{22}$ are necessary to characterize the VSI$_1$ PP-O null
radiation models.  Suppose that $\sigma = 0$; then either $\rho = 0$ and we have VSI, or $\rho \neq 0$ and $\Phi_{22} =
0$ which necessarily characterizes the vacuum VSI$_1$ models.

To show sufficiency, we assume $\kappa = \Phi_{12} = 0$ and then note that the remaining curvature components, $\Psi_4$
and $\Phi_{22}$, both have boost weight $-2$.  In the compacted (GHP) formalism \cite{penrid} the relevant operators
have boost weight $0$ or $1$ and the only spin coefficients with positive boost weight are $\sigma$ and $\rho$ with
weights $1$; it follows that the covariant derivative of either $\Psi_{ABCD}$ or $\Phi_{AB\dot{A}\dot{B}}$ will have
components with only negative boost weight. Therefore, all zeroth and first order curvature invariants vanish, implying
VSI$_1$.

\vspace{0.5cm}

(c) \textit{Petrov type O.}

The freedom in the frame can be used here to consider PP-type N and PP-type O null radiation separately, and it follows
trivially from the Bianchi identities that $\kappa = \sigma = \rho = 0$, so that we obtain VSI.  Therefore all Petrov
type O VSI$_0$  are VSI from the Bianchi identities.

\vspace{0.5cm}

In summary, the only spacetimes in the class VSI$_1$ that are not VSI  are of P-type N and all have $\kappa = \Phi_{12}
= 0$. The first of these VSI$_1$ models have $\sigma\Psi_4 = \rho\Phi_{22}$; exact solutions were found by
Pleba\'{n}ski \cite{kramer}.  The second of the VSI$_1$ models have $\sigma = \Phi_{22} = 0$, and these are the vacuum
Petrov type N solutions with $\rho = \Theta + i\omega \neq 0$.  If $\omega = 0$ these solutions belong to the
Robinson-Trautman class and all are known \cite{kramer}.  If $\omega \neq 0$ then the only twisting, vacuum, P-type N
solution known is that of Hauser \cite{kramer}.

There are other cases that may also be of interest. Notice the example in \cite{typIII} in which there are scalar
curvature invariants that are non-zero (constant -- depending on a cosmological constant) while all higher order scalar
curvature invariants are zero.

\section{$\varepsilon$-property}

A scalar invariant for a matrix is a polynomial of the
matrix entries that is invariant with respect to all changes of
basis.   It is easy to characterize all such invariants.  Let $M$ be
an $n\times n$ matrix. The characteristic polynomial of $M$ is given
by
$$p_M(x) = \det( x I - M) = x^n +\sum_{j=1}^n (-1)^j  \sigma_j(M)
x^{n-j}.$$ The expressions $\sigma_j(M)$ are called the elementary symmetric polynomials of $M$ and  are the scalar
invariants of $M$ ($\sigma_1(M)$ is just the trace of $M$ and $\sigma_n(M)$ is the determinant). All other scalar
invariants can be given as polynomials of $\sigma_1(M),\sigma_2(M),\ldots,\sigma_n(M)$. A matrix $M$ for which the
characteristic polynomial is  just $x^n$ is nilpotent. Now a matrix with the $\varepsilon$-property, i.e., the property
that all entries can be made smaller than every given $\varepsilon$ by  a change of basis, must be
nilpotent\cite{radjavi}. The converse is also true, that is, a nilpotent matrix must necessarily possess the
$\varepsilon$-property.
%
Therefore,  a matrix is VSI$_0$ if and only if it is nilpotent. Hence we anticipate that VSI spacetimes will have the
$\varepsilon$-property, and this is what we prove next.

{\bf Theorem:} For and only for VSI spacetimes (in arbitrary dimension $D$ and $C^{\infty}$ metric) one can find for
arbitrarily large $N$ and arbitrarily small $\varepsilon$ a tetrad in which all components of the Riemann tensor
and its derivatives up to order $N$ are smaller than $\varepsilon$. \\[2mm]
Proof: For non-VSI spacetimes there always exist a non-vanishing curvature invariant. Its value of course does not
depend on the choice of the tetrad and thus there does not exist a tetrad with the desired property. It was proven in
\cite{VSI} that in 4 dimensional VSI spacetimes the boost weight of all components of the Riemann tensor and its
derivatives is negative.  Thus with an appropriate boost we can make all components of the Riemann tensor and its
derivatives up to a desired order $N$ arbitrarily small \cite{arbD}. $\Box$ 

It was pointed out by Penrose in \cite{Penrose} that P-types III and N have "the property that gravitational density
can be made as small as we please by a suitable choice of time axis (following the wave)". It turns out that for
VSI$_0$ spacetimes, not only the gravity density but the energy-momentum tensor can be made arbitrarily small by an
appropriate boosting of the frame.  In the case of VSI$_1$ spacetimes we can also make the first derivatives of the
Riemann tensor essentially undetectable, and for VSI spacetimes it is possible to do this for arbitrarily large
derivatives as well. Since experiments measure tetrad components of the Riemann tensor and as every experiment has some
sensitivity limit, we can effectively, by an appropriate boost, "locally transform away" the Riemann tensor and its
derivatives.

It is of interest to consider if any of the VSI spacetimes satisfy the following stronger $\varepsilon$-property. We
shall say that the Riemann tensor has the uniform $\varepsilon$-property if, given an arbitrarily small $\varepsilon$,
there exists a tetrad in which the components of the Riemann tensor and all of its derivatives are smaller than
$\varepsilon$. Not all VSI spacetimes satisfy the uniform $\varepsilon$-property; this is shown by considering P-type N
vacuum VSI spacetimes with $\tau \neq 0$.  Let us denote

\begin{eqnarray}
X_{k}=C_{abcd;e_{1}\cdots e_{k}}n^{a}\overline{m}^{b}n^{c}\overline{m}^{d}m^{e_{1}}\cdots m^{e_{k}}, &\hspace{0.3in}
Y_{k}=C_{abcd;e_{1}\cdots e_{k}}\delta(n^{a}\overline{m}^{b}n^{c}\overline{m}^{d}m^{e_{1}}\cdots m^{e_{k}}).
\end{eqnarray}
By induction on $k$ we shall show that the component $C_{2424;3\cdots 3}=X_{k}=-k!\tau^{k}\Psi_{4}$ for all orders $k$.
From \cite{VSI} we have the following relations
\begin{eqnarray}
\kappa=\sigma=\rho=\epsilon=0, &\hspace{0.2in} \tau=\pi=2\beta=2\alpha, &\hspace{0.2in} \lambda=\mu=(2/3) \gamma,
\end{eqnarray}
where all of these spin coefficients are real, and $\nu$ is nonzero and complex as well. The Bianchi identities and NP
equations then give
\begin{equation}
\begin{array}{llll}
\delta \Psi_{4}=-\tau\Psi_{4}, &\ \ D \Psi_{4}=0, &\ \ \delta \tau = \tau^{2}, & \ \ D \tau =0.
\end{array}
\end{equation}
It can be shown directly that $X_{1}=C_{2424;3}=-\tau\Psi_{4}$, and using strong induction we assume that $X_{k}$ has
the required form. In general, the following recursive relation holds $X_{k}=\delta X_{k-1} - Y_{k-1}$, consequently
this implies that $Y_{k-1}=2(k-1)!\tau^{k}\Psi_{4}$.  Similarly, $X_{k+1}=\delta X_{k}-Y_{k}$, and on expanding $Y_{k}$
we observe that it is composed of terms with boost weight -2 and -1, but the boost weight -1 terms vanish as a result
of a similar proof found in \cite{VSI}.  To show this we note that in this case we have
\begin{equation}
\begin{array}{llll}
\thorn\Psi_{4}=0, &\ \ \thorn \tau=0, &\ \ \thorn \rho'=-2\tau^2=\thorn \sigma', &\ \ \thorn \kappa'=6\tau\rho'
\end{array}
\end{equation}
with commutators \cite{etapq}
\[\begin{array}{ll}
\thorn\edth-\edth\thorn=\tau\thorn\,, &\ \  \thorn\thorn'-\thorn'\thorn=2\tau(\edth+\edth')-(p+q)\tau^2.
\end{array}\]
Assuming that $\eta$ is a tetrad component of the Weyl tensor of arbitrary order $k$ with boost weight -2 such that
$\thorn\eta=0$, it is straightforward to show that the following boost weight -1 scalars,
\[\begin{array}{llllllll}
\thorn^{3}(\kappa'\eta), &\  \thorn^{2}(\sigma'\eta), &\ \thorn^{2}(\rho'\eta), &\ \thorn(\tau\eta), &\
\thorn(\tau'\eta), &\ \thorn\edth\eta, &\ \thorn\edth'\eta, &\ \thorn^{2}\thorn'\eta
\end{array}\]
all vanish. Therefore $Y_{k}$ consists of only the boost weight -2 term, hence we have that $Y_{k}=-2\tau X_{k}$ and
thus \mbox{$X_{k+1}=-(k+1)!\tau^{k+1}\Psi_{4}$.} Since the component $C_{2424;3\cdots 3}$ can be made arbitrarily large
by increasing the order, in this case the Riemann tensor cannot therefore satisfy the uniform $\varepsilon$-property.

A subclass of the VSI spacetimes for which the uniform $\varepsilon$-property is satisfied are those in which
$\nabla^{(N)}R_{a b c d}=0$, where $(N)$ denotes $N$ covariant derivatives.  Since only a finite set of components of
the Riemann tensor and its derivatives are nonzero, then by an appropriate boost all components of the Riemann tensor
and its derivatives can be made smaller than $\varepsilon$.  In the case of $N=1$ we have the VSI symmetric spaces in
which $\nabla_{e}R_{a b c d}=0$ (cases in which $N>1$ will be referred to as higher order symmetric spaces), we shall
show that this class is non-empty. We consider the following line-element

\begin{equation}
ds^2=2hdu^2+2dudv-dx^2-dy^2  \label{ppmet}
\end{equation}

\noindent and solve $\nabla_{e}R_{a b c d}=0$, assuming that $h=h(u,x,y)$. After an appropriate coordinate
transformation, which preserves the form of the metric, we find that $h=k(x^2+y^2)+c^2(x^2-y^2)$ where $k$ and $c$ are
arbitrary constants. Using the NP tetrad $\ell^{a}=\delta^{a}_{v}$, $n^{a}=\delta^{a}_{u}-h\delta^{a}_{v}$ and
$m^{a}=\left(i\delta^{a}_{x}-\delta^{a}_{y}\right)/\sqrt{2}$ it follows that the only non-vanishing spin coefficient is
$\nu$ with $\Phi_{22}$ and $\Psi_{4}$ being constants. If $k=0$ and $c\neq 0$ we recover the P-type N vacuum symmetric
space\cite{kramer}, if $k\neq 0$ and $c=0$ we obtain the P-type O, PP-type O null radiation symmetric
space\cite{kramer}.  These VSI symmetric spaces clearly satisfy the uniform $\varepsilon$-property. In P-type III it is
known that no symmetric spaces exist\cite{kramer}; however, the possibility remains that P-type III VSI spacetimes
satisfying the uniform $\varepsilon$-property may exist (for example, if $\nabla^{(N)}R_{a b c d}=0$ for $N>1$).

To illustrate a higher order symmetric space, consider (\ref{ppmet}) with $h=g(u)(x^2-y^2)$, a subclass of the P-type N
vacuum VSI spacetimes with $\tau=0$.  Next, apply a boost so that $l'=Al$ and $n'=A^{-1}n$ where the boost parameter
$A=Cg'(u)$ with $C$ constant.  Dropping the primes and working in the boosted frame we have the following non-vanishing
scalars, $\nu=-\sqrt{2}g(y+ix)/A^2$, $\gamma=A'/(2A^2)$ and $\Psi_{4}=-2g/A^2$.  It follows that the Weyl tensor has
the form \cite{bracket}

\begin{equation}
C_{a b c d}=\frac{1}{2}C_{2 i 2 j}\{\ell_{a}m^{(i)}_{b}\ell_{c}m^{(j)}_{d}\}  \label{gWeyl}
\end{equation}

\noindent where $i,j=3,4$, $m^{(3)}=\overline{m}$ and $m^{(4)}=m$, the only non-vanishing Weyl tetrad components are
$C_{2i2i}=2g/A^2$. 
Let $X_{0}=C_{2i2i}$ then (\ref{gWeyl}) is $C_{abcd}=\frac{1}{2}X_{0}\{\ell_{a}m^{(i)}_{b}\ell_{c}m^{(i)}_{d}\}$ and

\begin{equation}
\nabla_{e}C_{abcd}=\frac{1}{2}X_{1}\ell_{e}\{\ell_{a}m^{(i)}_{b}\ell_{c}m^{(i)}_{d}\}
\end{equation}

\noindent where $X_{1}=\Delta X_{0}+4\gamma X_{0}$.  It can be shown that the $n^{th}$ order covariant derivative of
the Weyl tensor has the following simple form,

\begin{equation}
\nabla_{e_{n}}\cdots\nabla_{e_{1}}C_{abcd}=\frac{1}{2}X_{n}\ell_{e_{n}}\cdots
\ell_{e_{1}}\{\ell_{a}m^{(i)}_{b}\ell_{c}m^{(i)}_{d}\}.                               \label{dnweyl}
\end{equation}

\noindent Proceeding inductively, we obtain the following recurrence relation $X_{n}=\Delta X_{n-1} + 2(n+1)\gamma
X_{n-1}$.
%
%
From (\ref{dnweyl}) we have that the only non-vanishing $n^{th}$ order tetrad components of the Weyl tensor will be
$C_{2i2i;2\cdots2}$. Again, by induction, one can show that $X_n=2A^{(n-1)}/(CA^{n+2})$ for all $n \geq 1$ (denoting
the $n-1$ derivative of A as $A^{(n-1)}$ and $A^{(0)}=A$). 
%
%
%
%
%
%

We now have an expression for the $n^{th}$ order derivatives of the tetrad components of the Weyl tensor

\begin{equation}
C_{2i2i;2\cdots 2} = \frac{2g^{(n)}}{(Cg')^{n+2}},  \label{bdnweyl}
\end{equation}

\noindent where it is assumed that $g' \neq 0$, otherwise the boost is degenerate. Therefore, for any $n \geq 2$ we can
obtain an $n^{th}$ order symmetric space simply by setting $g(u)$ to be any polynomial in $u$ of degree $n-1$. All of
these VSI spacetimes will satisfy the uniform $\varepsilon$-property; more generally this is also satisfied if there
exists a constant $M$ such that $|g^{(n)}| \leq M$ for all $n$ and $g' \neq 0$. On the other hand, we can use
(\ref{bdnweyl}) to find examples of VSI spacetimes that do not satisfy the uniform $\varepsilon$-property.  It is known
\cite{Stewart} that every geodesic of (\ref{ppmet}) is either of type 1 or type 2, where type 1 refers to geodesics in
the 2-surface $u$ and $v$ constant and type 2 refers to geodesics in the 2-surface $x$ and $y$ constant. Let us
consider type 2 geodesics, and set $x=x_0$, $y=y_0$.  We find that the tangent vectors are given by
$w^{a}=(a,b/(2a)-ag(u)(x_{0}^2-y_{0}^2),0,0)$ and parameterized by $u$.  Here, $\dot{u}=a$ is a constant and $b=1$ or
$0$ for timelike or null geodesics, respectively.  The NP tetrad defined above is parallelly propagated along such
geodesics, hence from (\ref{bdnweyl}) if the uniform $\varepsilon$-property is not satisfied at some order $k$ then we
obtain a parallelly propagated curvature singularity of order $k$.  That is, the curvature components of order $k$ in a
parallelly propagated frame become unbounded along the geodesic; when $k=0$ we recover the definition \cite{hawkell} of
a parallelly propagated curvature singularity.  In \cite{podobelan}, geodesic motion in vacuum Kundt type N solutions
with $\tau \neq 0$ have revealed the existence of parallelly propagated curvature singularities of order 0.

\section{Conclusion}

We have determined the necessary and sufficient conditions that characterize VSI$_1$ spacetimes.  Assuming VSI$_{0}$,
we have shown that in P-type III, VSI$_{1}$ implies VSI and in P-type O, VSI$_0$ implies VSI.  The only proper
VSI$_{1}$ spacetimes occur in P-type N and PP-type O with $\kappa=\Phi_{12}=0$.  In addition, the non-vacuum VSI$_{1}$
spacetimes are further characterized by $\sigma\Psi_{4}=\rho\Phi_{22}$, and the vacuum spacetimes have
$\sigma=\Phi_{22}=0$. It has been shown that the $\varepsilon$-property offers an alternative characterization of the
VSI spacetimes, in the sense that only for VSI spacetimes can a tetrad be found in which the Riemann tensor and its
derivatives up to any fixed order can be made arbitrarily small.  A strengthening of the $\varepsilon$-property leads
us to define the uniform $\varepsilon$-property; this condition determines a subclass of the VSI spacetimes where there
exists a tetrad in which the components of the Riemann tensor and all of its derivatives can be made arbitrarily small.
Some examples of VSI spacetimes satisfying the uniform $\varepsilon$-property have been presented.

\vspace{0.25cm}

\noindent{\bf Acknowledgements}

\noindent This work was supported, in part, by NSERC of Canada. VP was supported by GA\v{C}R-202/03/P017 and AP was
supported by KJB 1019403. NP would like to thank D. Pollney for the use of his program for computing invariants.
Portions of this work were made possible by the use of \emph{GRTensorII} \cite{grtensor} and Maple.


\begin{thebibliography}{}
\bibitem{arbD}
The generalization of this proof for arbitrary dimensions will be published in \cite{Dproof}.
\bibitem{bracket}
The \{\,\} operation used here is defined in \cite{chandra}.
\end{thebibliography}
\end{document}